\documentclass[pra,aps,showpacs,superscriptaddress,twocolumn,notitlepage]{revtex4-1}

\usepackage{amsmath,amsfonts,amssymb,amsthm,graphics,graphicx,epsfig,bbm}
\usepackage[colorlinks=true,citecolor=blue,linkcolor=blue,urlcolor=blue]{hyperref}
\usepackage[usenames]{color}
\usepackage{graphicx}
\usepackage{subfigure}
\usepackage{amsmath}
\usepackage{epsfig}
\usepackage{dcolumn}
\usepackage{bm}
\usepackage{color}
\usepackage{times}
\usepackage{epstopdf}
\usepackage{amssymb}
\usepackage{amstext}
\usepackage{latexsym}
\usepackage{hyperref}
\usepackage{amsfonts}
\usepackage{psfrag}
\usepackage{soul,xcolor}
\usepackage[normalem]{ulem}
\usepackage{dsfont}
\usepackage{adjustbox}

\newcommand{\beq}{\begin{equation}}
\newcommand{\eeq}{\end{equation}}
\newcommand{\beqa}{\begin{eqnarray}}
\newcommand{\eeqa}{\end{eqnarray}}

\usepackage{color}

\begin{document}

\title{Robust control of linear systems and shortcut to adiabaticity based on superoscillations}
\date{\today}

\author{Qi Zhang}
\affiliation{International Center of Quantum Artificial Intelligence for Science and Technology (QuArtist)
	and \\  Physics Department, Shanghai University, 200444 Shanghai, China} 
\affiliation{Laboratoire Collisions, Agr\'egats, R\'eactivit\'e, IRSAMC, Universit\'e de Toulouse, CNRS, UPS, France}

\author{Xi Chen}
\affiliation{Department of Physical Chemistry, University of the Basque Country UPV/EHU, Bilbao, Spain}
\affiliation{EHU Quantum Center, University of the Basque Country UPV/EHU, 48940 Leioa, Spain}

\author{David Gu\'ery-Odelin}
\affiliation{Laboratoire Collisions, Agr\'egats, R\'eactivit\'e, IRSAMC, Universit\'e de Toulouse, CNRS, UPS, France}

\begin{abstract}
With the advent of quantum technologies, control issues are becoming increasingly important.  In this article, we address the control in phase space under a global constraint provided by a minimal energy-like cost function and a local (in Fourier space) constraint resulting from a robustness criterion. The inverse engineering Lagrangian formalism developed here generalizes the one commonly used to describe the superoscillation phenomenon. It is  applied to both non-dissipative and dissipative quantum mechanics, and extended to stochastic thermodynamics. Interestingly, our approach also allows to improve the sensing capabilities by an appropriate control of the system.
\end{abstract}
\maketitle

Quantum control is nowadays  a crucial and implicit step in the manipulation of a wide variety of quantum systems, including, but not limited to, the quantum computation in the noisy intermediate-scale quantum  (NISQ) era \cite{EPJD,review2022}. Many methods have been developed in this field ranging from adiabatic control to optimal control theory \cite{EPJD,review2022} and shortcuts to adiabaticity \cite{STARMP}. For instance, protocols for high-fidelity fast transport or shuttling of charged ions \cite{Walther12,Bowler12,Qi2021} and neutral atoms  \cite{DGO08,davidmugapra} have been worked out, and are essential for the development of quantum technologies applications including quantum information processing \cite{Erikion,ErikBEC}, atom chip manipulation \cite{CorgierNJP,Becker,Amri} and metrology with cold atoms \cite{Dupont}. 

Besides reaching the desired target state, one cannot discard the question of the robustness of the designed protocols against various sources of noise, an inaccurate knowledge of the experimental parameters or model approximations \cite{Xipra11optimal,Xiaojingpra,an2015,QiJPB,Xiaopra18,EntropyDGO,Muga2022}. This concerns is particularly relevant to ions or neutral-atoms transport based on moving traps. Incidentally, such systems share the same dynamical equations as the load manipulation of mechanical cranes \cite{CranePRAppl,EnergyPRA}, and the motion of a charged particle in an external electric field \cite{OTCSTA}. These quantum-classical analogies allow to use
 control protocols in ubiquitous linear Lagrangian systems and compare the methods \cite{OTCSTA,controldavidsugny}. Furthermore, the above strategies can be applied to linear dissipative systems. This is illustrated with the motion of harmonically trapped Brownian particles driven by a time-varying force \cite{Cunuder16,TuPRE,JunPRR,New18}. 

In this article, we address the problem of the optimal control in a one-dimensional  (1D) phase space for a given protocol duration and under a global constraint provided by an energy cost and a robustness criterion that can be formulated as a local constraint in Fourier space. 
Such a control problem encompasses the transport and shuttling (see \cite{STARMP,Qi2021} and references therein) problems that have been addressed separately in the literature so far. The focus of the article is on the resolution of linear systems under such constraints, in the context of inverse engineering Lagrangian formalism. However, solving such a classical problem also provides the solution for the corresponding quantum problem \cite{davidmugapra}. We exemplify this idea with an application of our formalism to quantum transport described by coherent state obeying a Lindblad equation.

In the following, we first provide a short reminder on how optimal control theory is applied to linear systems. We then identify an intrinsic robustness against slight final time uncertainties. Later, we translate mathematically the requirement of robustness against an inaccurate knowledge of the strength of the moving transport potential, and explain how this extra constraint can be encapsulated in an inverse engineering Lagrange formulation.  Finally, the detailed applications are exemplified, ranging from classical and quantum mechanics to stochastic thermodynamics.

Consider a linear systems described by the following set of linear equations
$ \dot{\bm x} = \bm A \bm x(t) + \bm B \bm u(t)$,
where $\bm x(t) \in \mathbb{R}^n$ denotes the state vector and  $\bm  u(t) \in \mathbb{R}^q$ the control vector. The dynamics is assumed here to involve two time independent matrices: $\bm A \in \mathbb{R}^{n \times n}$ and $\bm B \in \mathbb{R}^{n \times q}$. We assume the system to be controllable, i.e.~it fulfills the Kalman's controllability criterion stating that the rank of the Kalman's controllability matrix $\bm K_c=[\bm B \; \bm A \bm B \; \bm A^2 \bm B \; ... \; \bm A^{n-1} \bm B]$ is $n$, see \cite{controldavidsugny}. The  solution then reads 
$\bm x(t) = e^{\bm A t} \bm x_0 +\bm I_u(t)$ with $\bm I_u(t)=\int_{0}^{t} e^{\bm A (t-t') } \bm B \bm u(t') dt' $.
The control problem consists in designing the time-dependent control vector to drive the system from an initial state ($t_0, \bm x_0$) to the desired target state ($t_f, \bm x_f$) in a time interval that lasts $t_f$. The boundary condition at final time imposes $\bm I_u(t_f)=\bm x_f -  e^{\bm A t_f} \bm x_0$. In most physical situations, extra requirements are needed such as the minimization of a cost function $E=\int^{t_f}_{0}  f[\bm x(t), \bm u(t), t] dt$ that depends on both the trajectory $\bm x(t) $ and the driving $\bm u $. such a minimization problem can be solved thanks to the Lagrangian multiplier formalism:
\begin{equation}
\label{eqEL}
\frac{\partial E}{\partial \bm u} - \bm\mu_0^T\frac{\partial  \bm G_0 (\bm u)}{\partial \bm u} =0 ,
\end{equation} 
where $ \bm G_0 (\bm u)=\bm I_u(t_f)-\bm x_f + e^{\bm A t_f} \bm x_0$.
For the cost function, $E_1=\int_{0}^{t_f} \bm u^T \bm udt$ the Lagrangian multiplier is given by applying the controllability Gramian matrix $\bm\mu_0=\bm W (t_f)^{-1}\big(\bm x_f - e^{\bm A t_f} \bm x_0 \big)$ \cite{Gramian}. Alternatively, the very same results can be recovered using the Pontryagin formalism of optimal control theory. 

For a particle in a moving harmonic trap characterized by its time-dependent center at $x_0(t)$ and its constant angular frequency $\omega_0$,   the dynamical equation reads $\ddot x + \omega_0^2x=\omega_0^2x_0(t)\equiv u(t)$, where $u(t)$ is the control parameter that drives the system.
This second-order differential equation is readily transformed into a set of two first order differential equation with the variable $x$ and $y=\dot x$.
The control problem we focus on consists in designing the time-dependent control parameter $u(t)$ to transfer the system from an initial condition $(x_i=0, ~y_i=0)$ to
an arbitrary target $(x_f=r \cos \varphi,~y_f= r \omega_0\sin \varphi)$ in phase space  in a finite amount of time $t_f$.
For $y_f\neq 0$, such a control problem is referred to as shuttling \cite{Erikion,Qi2021}. The particle acquires the desired velocity at a given position after application of a proper driving. For a vanishing final velocity $y_f=0$, this problem is nothing but a transport problem for which the particle initially at rest is displaced from $x=0$ to $x=x_f$ and remains at rest afterwards \cite{DGO08,Erikion}.

\begin{figure}[t]
\begin{center}
\scalebox{0.55}[0.55]{\includegraphics{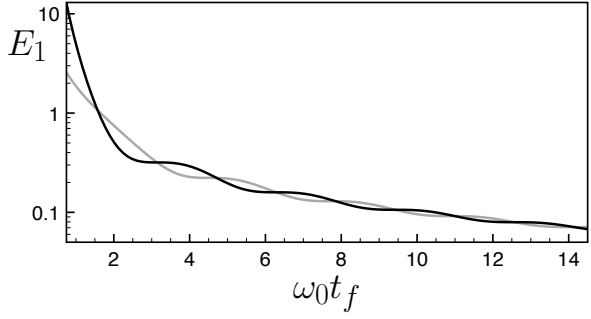}}
\caption{The energy cost $E_1$ as a function of the protocol duration $t_f$ in units of $\omega_0^{-1}$ for transport ($\varphi=0$, solid black line) and shuttling ($\varphi=\pi/2$, solid gray line). Parameters: $d=1$, $\omega_0 t_f=10$.}
\label{figE1}
\end{center}
\end{figure}

Using the previous formalism for the minimization of the standard cost function $E_1=\int_0^{t_f} u^2 dt$, the 2-dimensional Lagrangian multiplier $\bm \mu _0 $ can be readily derived from the boundary conditions yielding the following expression for the control parameter
$$
u_1(t)\!=\!\frac{2 r\omega_0^2 [ \varphi_f \sin (\varphi\!+\!\varphi_f\!-\!\omega_0t)\! -\! \sin (\omega_0t \!+\! \varphi)\sin (\varphi_f) ]}{\varphi_f^2 \!-\!\sin ^2(\varphi_f)},
$$
with $\varphi_f= \omega_0t_{f} $.
The corresponding minimal energy cost $E_1$ is
\begin{equation}
\label{eqE1so}
E_1 = 2 r^2 \omega_0^3 \frac{\varphi_f + \cos(2 \varphi + \varphi_f) \sin(\varphi_f)}{\varphi_f^2 -\sin^2 (\varphi_f)}.
\end{equation}
This energy cost provides the lower bound to reach the target ($r \cos \varphi, \, r \omega_0\sin \varphi$) in an amount of time $t_f$. In Fig.~\ref{figE1} (a), we have depicted this energy as a function of the final time for both a transport problem (target with $\varphi=0$, black solid line) and a shuttling process (target with $\varphi=\pi/2$, gray solid line). As intuitively expected, the shorter the driving time $t_f$, the larger the energetic cost $E_1$, and $E_1$ is a decreasing function of $\omega_0t_f$. Interestingly, this result reveals the existence of plateaus for which the change of $t_f$ has a negligible impact on the energy cost, meaning that the corresponding solution are intrinsically robust against a small variation of the protocol duration $t_f$ for a succession of windows of final time. The energetic cost (\ref{eqE1so}) is the lowest possible as it is derived without any constraints. We denote by $E_1^{\rm (min)}$ this lower bound. 

Next, we add extra constraints and derive an inverse engineering Lagrangian approach to build protocols robust against the exact knowledge of the trap frequency $\omega_0$. The energetic cost $E_1$ associated to such a requirement will be quantified and compared to $E_1^{\rm (min)}$.
As a first example, we elaborate the transport problem, i.e. a target $r=d$ and $\varphi=0$ to be reached in an amount of time $t_f$. For an arbitrary driving $x_0(t)=u(t)/\omega_0$, the final excitation energy reads \cite{davidmugapra}
\begin{equation}
\Delta E (\omega_0;t_f) = d^2\omega_0^2 \left|\int_0^{t_f} \ddot{x}_0 (t)e^{-i \omega_0 t}dt \right|^2 \equiv  \left|  g (\omega_0;t_f)  \right|^2.
\label{defg}
\end{equation}
A perfect transport (without residual ossilations) requires $\Delta E (\omega_0;t_f) =0$. A transport robust against the exact value of $\omega_0$ requires to smoothen $\Delta E$ as a function of $\omega$ about $\omega_0$ \cite{davidmugapra}:
\begin{equation}
\frac{\partial^p \Delta E (\omega;t_f)}{\partial \omega^p}\bigg|_{\omega_0}=0, \;\; \mbox{with}\;\; p \geq 1.
\label{requir1}
\end{equation}
The calculation associated to this robustness issue are made simpler by using the complex variable $z(t)=\dot x/\omega_0 +i x$. The equations of motion are then $\dot{z}(t)=i \omega_0 z(t) +u(t)/\omega_0$. Under the driving $u(t)$, the complex solution is $z(t)= e^{i \omega_0 t} z(0) +\tilde{u} (\omega_0)/  \omega_0$ where $\tilde{u} (\omega_0) = \int^{t_f}_0 u(t) e^{-i \omega_0 t} dt \equiv a_0$. For an arbitrary target $z(t_f) = r( \sin \varphi +i \cos \varphi)$, $a_0=r\omega_0\left( \sin \varphi +i \cos \varphi \right) e^{-i\omega_0 t_f}$.

For transport, the driving function obeys the boundary conditions: $u(0) = 0$, $u(t_f)=\omega_0 d$, and $\dot{u}(0) =  \dot{u}(t_f)=0$.
To fulfill the robustness requirements Eq. (\ref{requir1}), we impose the nullity of the successive derivative of  $g (\omega;t_f)$ (defined in Eq.~(\ref{defg})) with respect to $\omega$ about $\omega_0$. We infer the following set of extra constraints on the Fourier function  $\tilde{u} (\omega)$ and its successive derivatives:
\begin{equation}
\omega_0\tilde{u}^{(p)}(\omega_0)+ 2p \tilde{u}^{(p-1)}(\omega_0) + \frac{p(p-1)}{\omega_0}\tilde{u}^{(p-2)}(\omega_0) = \lambda,
\end{equation}
where $\lambda=d(\varphi_f+ip)(-it_f)^{p-1}e^{-i\varphi_f}$. We find $\tilde{u}^{(m)}(\omega_0)=a_m$ with $a_1=d(\varphi_f-i)e^{-i\varphi_f}/\omega_0$, $a_2= d(-i\varphi_f^2-2\varphi_f+2i)e^{-i\varphi_f}/\omega_0^2$ and so on.

More generally, the class of problem to be solved here requires the minimization of $E=\int_{\cal D}  f(t) f^{*}(t) \mathrm{d}t$ on a compact support  ${\cal D}$ under a discrete number of constraints on the Fourier transform of $f$ and its derivatives for a finite set of angular frequencies $\omega_j$: $
\tilde{f}^{(m)}(\omega_j) = \int_{\cal D} (-it)^m f(t) e^{-i \omega_j t} \mathrm{d}t=\tilde f^{(m)}_j$. For a single constraint on the Fourier transform, such a problem has been solved in the context of superoscillations detailed in \cite{aharonove,berry1,berry2,SuperOsc}. We generalize here this approach to enforce the robustness requirement.
The solution relies on the Euler-Lagrange formalism:
\begin{equation}
\frac{\delta E}{\delta f} +\sum_j \sum_{m=0}^p\nu_{j,m} \frac{\delta \tilde{f} ^{(m)}(\omega_j)}{\delta f} =0,
\end{equation}
where $\nu_{j,m}$ are Lagrange multipliers. Interestingly, this set of equations can be directly solved by a simple polynomial ansatz
\begin{equation}
\label{eqfx}
f(t) = \sum_j \bigg( \nu_{j0}^* + \sum_m (i t)^m ~\nu_{j,m}^* \bigg) e^{i \omega_j t}.
\end{equation}
The Lagrange multipliers, $\nu_{j,m}$, are subsequently determined by the constraints. The inverse engineering procedure amounts here to fulfilling the boundaries condition to set up the driving. One should be careful to keep $u(t)$ real as it coincides with the position of the trap in real space. To get a real driving function, we simply exploit the linearity and define the control function as $u(t)=(f(t)+f^*(t))/2$.

\begin{figure}[t]
\begin{center}
\includegraphics[height=9.cm,angle=0]{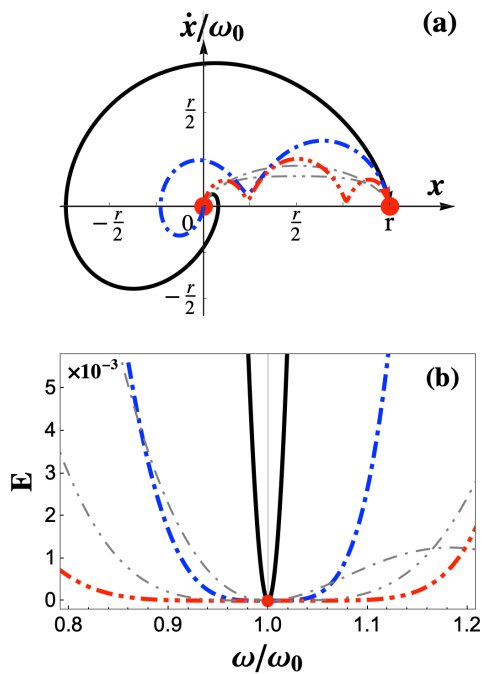}
\caption{In the case of transport ($\varphi=0$), (a) trajectories in phase space, and (b) the final excitation energy as a function of $\omega$  are compared between (i) Fourier Euler-Lagrange method with no extra robust constraint protocol (black solid line), $p=1$ robust control (blue dot-dashed line) and $p=2$ robust one (red double-dot-dashed line) and (ii) and Fourier method for $N=1$ one frequency robust protocol (gray dot-dashed thin line) and $N=2$ two frequencies robust one (gray double-dot-dashed thin line). Parameters: $d=1$, $\omega_0 t_f=10$.}
\label{figtrans}
\end{center}
\end{figure}

In Fig.~\ref{figtrans}, we plot the excess of energy after transport in a harmonic potential as a function of its angular frequency $\omega$ for protocols optimized for a given $\omega_0$ and for a transport duration $\omega_0t_f=10$. We have computed the solutions under the robustness requirements on the first (blue dot-dashed line) and second (red double-dot-dashed line) derivative of the Fourier transform of the driving function. We clearly observe a strong improvement associated to the local flatness about $\omega_0$ compared to the solution obtained by optimal control without constraints (black solid line). 
We have also compared the solutions, see the gray lines in Fig. \ref{figtrans}, obtained by the so-called Fourier method \cite{davidmugapra} that has been developed to address similar requirements. 

In fact, the formalism developed here is well-adapted to the optimization of other class of transport problems, since solving this classical problem amounts to solving its quantum counterpart as explained in Ref.~\cite{davidmugapra}. The transport problem in the presence of dissipation requires to solve the equation of motion for the density matrix.
Regarding transport with losses, the dynamics of coherent state $\left.| \alpha \right\rangle$ is given by the following master equation:
\begin{equation}
\label{eq-coherent}
\frac{d \rho }{dt}= -\frac{i}{\hbar}[H, \rho]+ \frac{\Gamma}{2} (2a \rho a^{\dag}- a^{\dag} a \rho-\rho a^{\dag} a),
\end{equation}
where $\Gamma$ is the Lindblad operator and $H$ the Hamiltonian, $H=p^2/2m+m \omega^2 [\hat{x}-x_0(t)]^2/2$, with $x_0(t)$ being the center of moving potential \cite{Erikion,DGO08}. This problem can be solved exactly whatever is the driving using a time-dependent coherent state $\rho(t)=\left.| \alpha \right\rangle \left\langle \alpha \right|$ with
$\alpha(t)= (i\omega/\sqrt{2} a_0 ) e^{-(i \omega+\Gamma/2)t}\int_0^t e^{(i \omega+\Gamma/2)t'} x_0(t')dt'
$.  This problem belongs to the same class of problems as previously studied since it can be expressed as a linear system on the variables $x_1=\left\langle \hat x \right\rangle$, $x_2=d \left\langle \hat x \right\rangle/dt$: $\dot x_1 = x_2$ and $\dot x_2=- (\omega_0^2 + \Gamma^2/4)x_1-\Gamma x_2+\omega_0^2 x_0(t)$.

\begin{figure}[t]
	\begin{center}
		\includegraphics[height=9.cm,angle=0]{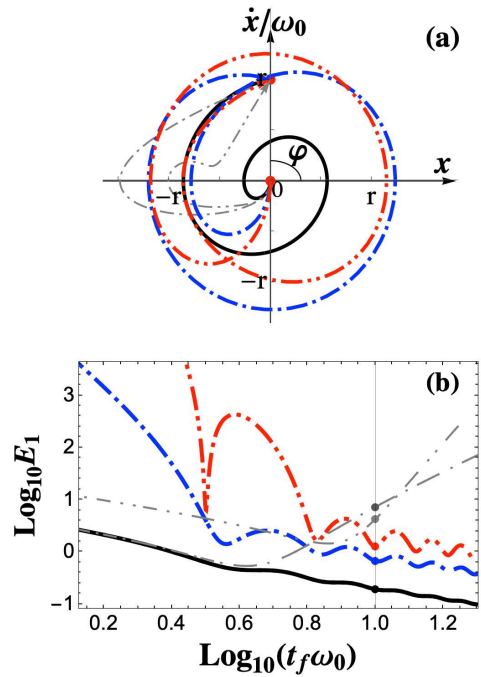}
	\end{center}\caption{ Shuttling protocols (target $\varphi=\pi/2$). (a) Phase space trajectories associated to the different protocols performed over a $t_f=10/\omega_0$ time interval. (b) Energy cost of the different protocols as a function of the protocol duration. Same notations as Fig.~\ref{figtrans}.}
	\label{figshuttling}
\end{figure}

Moreover, the transport of a mesoscopic object in stochastic thermodynamics can also be captured and optimized by
our formalism. For this purpose, it turns out to be more convenient to work with the distribution function rather than with the Langevin-like stochastic differential equations.  In the overdamped regime, the distribution function $\rho (x,t)$ obeys the following Fokker-Planck equation:
\begin{equation}
\label{eq-overFP}
\gamma\partial_t \rho (x,t) = \partial_x[ m\omega_0^2\big(x-x_0(t) \big) \rho ] + \gamma D \partial_{xx}^2 \rho,
\end{equation}
where $\gamma$ is the friction coefficient and  $D=k_B T/\gamma$ the diffusion coefficient.
The solution of the FP equation is provided by a Gaussian $\rho (x,t)=\sqrt{\frac{\alpha}{\pi}} \exp[ -\alpha (x-x_c(t) )^2 ]$ with $\alpha=2\gamma D/(m \omega_0^2)$. The time-dependence of $\rho (x,t)$ is encapsulated in the variable $x_c(t)$ governed by the linear system $\dot x_c + (m\omega_0^2/\gamma) (x_c-x_0) =0$ for which the previous formalism applies.
These results can be readily generalized to the underdamped regime. For this purpose, we introduce the phase space distribution $\rho(x, v, t) $ that obeys the Kramers equations \cite{Cunuder16}.
A Gaussian solution can be readily worked out $\rho(x, v, t) = \mathcal{N} \exp[ -\alpha (x-x_c(t) )^2
-\beta (v -v_c(t))^2 ]$ with $\alpha = m\omega_0^2/2k_B T$ and $\beta =m/2k_B T \gamma$, and with a time variation for the parameters $x_c$ and $v_c$ dictated by the following set of linear equations: $\dot x_c= v_c $,  $\dot v_c = - \omega_0^2 x_c - \gamma v_c/m + 
\omega_0^2 x_0(t)$.

Besides targets in phase space located on the position axis, our protocol encompasses the shuttling problem for which the particle acquire a finite velocity. The results obtained for such shuttling protocols are summarized in Fig.~\ref{figshuttling}.
As a general result, the inverse engineering Lagrangian method developed to ensure robustness provides systematically a much lower $E_1$ energy for a time duration of the protocol larger than $t_f>10/\omega_0$ compared to the Fourier method  \cite{davidmugapra}. The latter turns out to be more efficient for extremely short shuttling protocol duration but at the expense of a large energy cost $E_1$. In Fig.~\ref{figshuttling}, we also provides for comparison the minimum energy obtained by optimal control theory for the minimization of the energy cost in the absence of any other constraints (black solid line). We therefore clearly identify the minimum cost of the robustness requirement.

\begin{figure}[t]
	\begin{center}
		\includegraphics[width=\columnwidth]{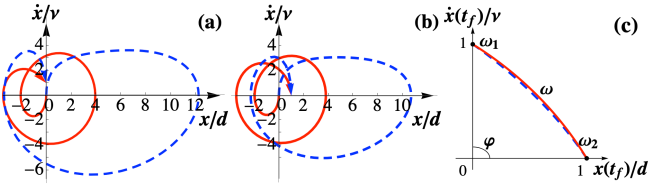}
	\end{center}\caption{Sensing the frequency between $\omega_2=1.05\omega_1$ by separating the corresponding target states as follows: $(r_1=\nu, \varphi_1=\pi/2)$ and $(r_2=d, \varphi_2=0)$. The comparison is between the following two protocols: Fourier Euler-Lagrange approach (red solid line) and Fourier method (blue dashed line). Parameter: $t_f=10$ and $\omega_1=1$.}
	\label{figsensing}
\end{figure}

Finally, the inverse engineering Lagrangian formalism enables one to design a driven protocol for quantum sensing, with the objective to measure the trap frequency or its drift as a function of time. To this end, a protocol is designed to  assign  different targets to two close but different trapping frequencies $\omega_1$ and $\omega_2=(1+\epsilon)\omega_1$ ($\epsilon\neq 0$). Such a strategy is reminiscent of the fingerprinting method used in NMR \cite{Ma13,Ansel17}. 
We therefore consider here the same control function that drives two independent set of equations: $\dot x_j=y_j$ and $\dot y_j=-\omega_j^2y_j+u(t)$. To transfer the system from the initial state $\big(x_j(0), \dot x_j (0) \big)=(0, 0)$  to distinguishable target states $\big( x_j(t_f), \dot x_j (t_f) \big) =(r_j \cos \varphi_j, r_j \omega_j \sin \varphi_j)$. 
The Fourier Transform of control $u(t)$ should be equal to $
\tilde{u} (\omega_j) = \omega_j r_j\left( \sin \varphi_j +i \cos \varphi_j \right) e^{-i\omega_j t_f}$. Such conditions are readily fulfilled with
an ansatz of the form $u(t) = \sum_{j} [ a_j \cos(\omega_j t) -  b_j \sin(\omega_j t) ]$.
For example, the two frequencies case refers to $j=1,2$ and the target can be $r_1=\nu, \varphi_1=\pi/2$, $r_2=d$ and $\varphi_2=0$.
Such conditions are readily fulfilled with 
an ansatz of the form $u(t) = \sum_{j=1}^2 [ a_m \cos(\omega_m t) -  b_m \sin(\omega_m t) ]$ . As shown in Fig.~\ref{figsensing}, we plot in red the evolution of the trajectories for the two different $\omega_i$ towards their target, along with the final state in phase space $(x_f, \dot x_f)$ for different trapping angular frequencies $\omega \in (\omega_1, \omega_2)$. It demonstrates the possibility to magnify the sensitivity. Indeed, in the absence of driving, the information on the trapping angular frequency cannot be recovered as the system remains at rest. This sensing protocol is particularly relevant when the time duration of the protocol $t_f$ remains smaller than the natural timescale $\pi/(2 \varepsilon \omega_1)$ which would generate an angle separation in phase space equal to $\pi/4$ for a step driving (sudden switch of the driving function to a constant value). This is  the reason why in Fig.~\ref{figsensing}, we have chosen the parameters to fulfill the inequality $\omega_1t_f \ll \pi/(2 \varepsilon)$.  The curves in blue dashed lines have been obtained using de Fourier method \cite{davidmugapra}. The same conclusion holds concerning the sensing protocol. It is worth noticing that the Lagrange method limits considerably the extension of the trajectory to obtain a similar sensitivity. This can be a strong asset to limit the role of anharmonicities in such a protocol \cite{an2015,QiJPB}. The fingerprint sensing method discussed here can be readily generalized to a larger number of $\omega_i$.

To conclude, we have developed a formalism inspired by Lagrangian minimization to tackle optimal control of linear systems taking into account the different constraints, such as energy-like minimization and robustness criterion. Our inverse engineering  Lagrangian formalism generalizes
the one for driving quantum systems with superoscillations to enforce the robustness or sensitivity requirements. This has been adopted to both non-dissipative and dissipative linear systems, describing 
the transport of ions or neutral atoms with/without loss, and a mesoscopic object in stochastic thermodynamics. Our results pave the way for the design of optimal shortcuts to adiabaticity in a class of general linear systems with broad applications in quantum technologies.

This work has been financially supported by NSFC (12075145), STCSM (2019SHZDZX01-ZX04), the Agence Nationale de la Recherche through grant ANR-18-CE30-0013, EU FET Open Grant EPIQUS (899368),  QUANTEK project (KK-2021/00070). X.C. acknowledges the Ram\'on y Cajal program (RYC-2017-22482).

%
%
%



\begin{thebibliography}{99}
\bibitem{Gramian} The controllability Gramian matrix is defined by  $\bm W(t_f)=\int_{t_0}^{t_f} e^{\bm A(t_f-t')} \bm B \bm B^T  e^{\bm A^T(t_f-t')} dt' $.
\end{thebibliography}
\end{document}